\documentstyle[aps,prl,epsf,epsfig,floats]{revtex}
\tightenlines
\begin{document}
\draft
\title{Kondo Effect and Persistent Currents in a Mesoscopic
Ring:\\Numerically Exact Results.}
\author{E.V. Anda$^1$, C. Busser$^2$, G. Chiappe$^2$ and M. A. 
Davidovich$^1$} 
\address{\it $^1$Departamento de F\'{\i}sica, Pontificia Universidade 
Cat\'olica do Rio de Janeiro, C.P. 38071-970, Rio de Janeiro, RJ, Brazil \\
\it $^2$Departamento de F\'{\i}sica,
Universidad de Buenos Aires CP 1428-Nu\~nez, Buenos Aires, Argentina}

\maketitle
\begin{abstract}
We study the persistent current circulating along a mesoscopic ring with a 
dot side-coupled to it when threaded by a magnetic field. A cluster including 
the dot and its vicinity is diagonalized and embedded into the rest of the 
system. The result is numerically exact. We show that a ring of any size can be
in the Kondo regime, although for small sizes it depends
upon the magnetic flux.
In the Kondo regime, the current can be a smooth or a strongly dependent
function of the gate potential according to the structure of occupation of the 
highest energetic electrons of the system.
\end{abstract}
\date{}

\twocolumn[\hsize\textwidth\columnwidth\hsize\csname@twocolumnfalse\endcsname]
%\twocolumn

%\documentstyle[preprint,aps,prl,epsf]{revtex}

%\draft

In recent years a great effort has been dedicated to the understanding of 
resonant tunneling taking place in a QD connected to two leads
\cite{1,6}. In theses systems it has been experimentally and theoretically
established 
\cite{1,4,5} that under certain conditions, at very low
temperatures, when the dot is charged and possesses a net spin, it couples 
antiferromagneticaly to the conducting electron spins giving rise to a Kondo 
state. This is reflected in the transport properties of this system, 
because the Kondo effect maintains the dot at resonance providing a channel at
the Fermi level through which the electrons can tunnel the dot. In general
the device is designed such that the current goes through the dot itself. 

Very recently  some studies have proposed another configuration in which the 
dot is laterally linked to a quantum wire \cite{4,5}. This situation mimics
in some extent a metallic compound when doped by magnetic impurities 
\cite{7}. In this case the localized level of the magnetic impurity, a $f$ or
$d$ orbital, is connected laterally to the conduction band, without 
participating in the path through which the electrons go along the system. 
Similarly to the problem of the metal doped with magnetic impurities, 
theoretical studies of a dot laterally attached to a wire have shown that 
the Kondo effect interferes with the transport channel reducing and eventually 
eliminating the transmission of charge along it \cite{8}. A similar
configuration has been proposed where the dot is laterally attached to a
ring threaded by a magnetic field, which creates a circulating persistent 
current (PC) \cite{6}. In this case however, contrary to the previous one,
theoretical 
studies \cite{9,10} have predicted that, in the Kondo regime, the current
is not affected by the dot. The influence of the dot is recovered when the
temperature is greater than the Kondo temperature  $ T_k $, or by reducing
the size of the ring such that  $T_k \ll \Delta $, where $\Delta$ is the ring level spacing.
This last limit is believed to be a situation where the Kondo resonance is 
strongly suppressed\cite{9,10}. However, it has been shown that the behaviour results to 
be different when the geometry of the system is such that the electron goes 
through the dot in order to encircle the magnetic flux. In this case the limit
$T_k \ll \Delta $  does not eliminate the Kondo effect \cite{11}. 

In this letter we study, in an exact numerical way, the persistent current of 
a ring encircling a magnetic flux coupled to an adjacent QD in the Kondo regime.
Our purpose is to contribute to the clarification of some fundamental aspects 
of this problem.

This study permits us to conclude that when the dot is in the Kondo regime,  the
 ring states near the Fermi level are not conducting, as it is 
the case of the wire states, showing that there is no contradiction between the behaviour of 
these two systems. However, while in the wire this effect quenches the current, 
for the ring, the states that are below the Fermi level by an amount grater than $T_k$ 
still contribute to the PC. There is no destructive interference of the current
produced by a dot laterally attached to a ring, as in the case the dot is
attached to a wire. The influence of the dot on the PC, in the Kondo regime, 
depends upon the number of electrons $N_e$ and, for small rings, on the magnetic
flux $\phi$, that will be considered in units of $\phi_0$, the quantum of flux.
 For the case in which $N_e = 4n$, where $n$ is an arbitrary integer, and 
$\phi=0.5$  or, $N_e=4n+2$ and $\phi=0$, the Kondo effect reduces the influence
of the dot on the current when, by changing the gate potential, an
electron goes from the ring to the empty dot. In the limit $U =\infty$, 
previous calculations \cite{9,10} have predicted a complete independence of 
the PC on the gate voltage. In this limit, we agree with these results for 
$N_e=4n,\phi=0.5$ and $N_e=4n+2, \phi=0$.
However, this is not the case 
 for the opposite situation, $N_e=4n,\phi=0$ or $N_e=
4n+2,\phi=0.5$, where it is the entrance of a hole to the filled dot, which maintains 
the current invariant. Moreover, the cases $ N_e = 4n+1$ and $N_e=4n+3$ 
correspond to a completely different behaviour, with a strong dependence 
of the PC on $ V_g$, as discussed below.

Another important point to be clarified is the physical process that takes 
place when the length of the ring is reduced, ($\Delta \gg T_k$). The
Kondo 
effect is not necessarily eliminated in this case \cite{12} and can even be 
enhanced as the Kondo temperature does not have an exponential dependence 
upon $\epsilon_d/\Gamma$,where $\epsilon_d$ is the energy level of the dot and
$\Gamma=t'^{2}/W$, where $t'$ is the hopping matrix element between the dot and
the ring and $W$ is the band width. The crucial point here is related to the 
configuration of the most energetic occupied states that depends upon $N_e$ and 
$\phi$.

The properties of the system are calculated using a Lanczos algorithm to 
obtain the ground state of a cluster containing the dot that is afterwards 
embedded into the rest of the system \cite{11,13}. As a consequence of the
procedure used, the transport properties result to converge very rapidly to 
its exact value increasing the size of the cluster\cite{13}.

The system is represented by an Anderson impurity first-neighbour tight
binding 
Hamiltonian of $N$ sites. In the cluster the QD state is denoted by $d$
and the other sites 
are numbered  from $r$ to $\bar r$. The total Hamiltonian can be written as 
the sum of two terms representing the cluster ($H_c$) and the rest of the
ring and the coupling between them ($H_r$). Measuring the energies in units
of the matrix hopping elements, $H_r$ can be written as,
\begin{eqnarray}
\nonumber
H_r=\sum_{<ij>\atop\sigma}c^\dagger_{i\sigma}c_{j\sigma}+  
e^{i\pi\phi}\sum_\sigma (c^\dagger_{r\sigma}c_{r+1\sigma}+c^\dagger_{\bar
r-1\sigma
}c_{\bar r\sigma})+ c.c.
\end{eqnarray}
where $i$ and $j$ are the first neighbours atomic sites outside the cluster, 
belonging to a chain of $ L $ sites. The Hamiltonian of the cluster with 
$N-L$ sites, $H_c$, is
given by,
\begin{eqnarray}
\nonumber
H_c=\sum_\sigma\left[(V_g +{U \over 2} n_{d\bar\sigma})n_{d\sigma} 
+t'c^\dagger_{d\sigma}c_{o\sigma}+c.c+\sum_{m,n} c^\dagger_{m\sigma}
c_{n\sigma}\right]
\end{eqnarray}
where $U$ corresponds to the electronic repulsion at the dot. The hopping 
matrix element between the dot state and its nearest neighbour is given by 
$t'$ and $m$ and $n$ stand for the atomic sites of the ring belonging to the cluster.

The persistent current is calculated by solving a Dyson equation for the 
dressed Green function $\hat G=\hat g+ \hat g\hat T\hat G$ where $\hat g$
is the cluster Green function matrix  obtained by the Lanczos method and
$\hat T$ is the matrix representing the coupling Hamiltonian between the
cluster and the rest of the system. To take into account charge
fluctuation, we write $\hat g$ as a combination of the Green function of
$n$ and $n+1$ particles as described in \cite{13}. Then the PC is
calculated as ${1\over2\pi} \int (G_{i,i+1}(w)-G_{i-1,i}(w))dw=\int I(w)dw$.

The transport properties are studied in the $\Delta{>\atop <}T_k$ regimes. We 
adopt a set of parameters compatible with the experimental systems used to 
detect the Kondo effect, $t'^{2}/WU=0.01$ \cite{3}. We analyze the
case in which the system is supposed to be isolated with a constant number
of electrons $N_e$, independent of the gate potential applied to the dot. 

Consider first a small ring with $N$ sites such that $\Delta>T_k$. We show that the only 
relevant states that control the physics of the problem are the dot level and the two upper 
occupied states of the ring. These two ring levels have the same energy if the external 
flux is $\phi= 0.5$, ($\phi=0$), corresponding to the ring wave vectors 
$(k_n={2\pi\over N}(n+\phi), k_{-n-1})$, $(k_n,k_{-n})$, 
or acquire its maximum splitting, $\Delta\sim W/N$, when $\phi=0.25$. 
These two states can be occupied by 4, 3, 2, or 1 electrons. 
These four situations, $(N_e = 4n+3; 4n+2; 4n+1; 4n)$, 
correspond to different ways in which the 
PC is created when an external magnetic flux is applied to the ring \cite{14}.
In order to illuminate the problem at hand, we study a system with $N=N_e=3$.
The three sites represent the last two occupied states of the ring and the dot state 
with energies $\epsilon_1$,$\epsilon_2$ and $\epsilon_d$, respectively. Although these 
three states could describe only a very small ring, they determine as well crucial 
aspects of the physics of a ring of any size, as we discuss below.
 Using perturbation theory it is possible to 
obtain an algebraic expression for the Kondo temperature by calculating the difference of  
energy between the Kondo ground state and the first excited state. 
Taking $\epsilon_2=0$,$\epsilon_d<0$ and $\epsilon_1>0$, the Kondo temperature can be written 
as  $T_k=t'^{2}((\sqrt(2)+1/2) \epsilon_d+\epsilon_1)/\epsilon_d
(\epsilon_1-\epsilon_d
)$.We mimic the the effect of the flux  changing the value of  $\epsilon_1$ through the relationship
$\epsilon_1 = {\pi\hbar^2N_e\over mN^2}(0.5-\phi)$.
 When $\epsilon_1=-(\sqrt(2)+1/2)\epsilon_d$ the Kondo temperature is
$T_k=0$ and 
it is maximum when $\epsilon_1=0$. As a consequence, for small rings,
when $\Delta \gg T_k$, the system is in the Kondo regime if the flux
satisfies $0.5> \phi> 0.5+(\sqrt(2)+1/2)\epsilon_d mN^2/\pi\hbar^2 N_e$.
This result is
reflected 
in the spin-spin correlation between the dot spin $S_d$ and the spins $S_1$ 
and $S_2$ corresponding to the other two levels. As shown in fig. $1a$, 
these correlations are maxima in the vicinity of $\epsilon_1= 0$, $ (\phi=0.5)$, while are 
negligible for large values of $\epsilon_1$, $(\phi=0.25)$, indicating the absence of a 
Kondo ground state in this case. We learn from this study that, depending upon the magnetic flux, a 
QD side-coupled to a small ring, $(\Delta > T_k)$, can be in the Kondo regime
 with a $T_k$ that involves the
energies of the system in an algebraic non-exponential way \cite{12}. 
This is consistent with the enhancement of the persistent current going through a 
dot and encircling a ring, when the system is in the Kondo regime as 
reported in \cite{11}.

To study the problem for a small ring we concentrate our
analysis to the neighbourhood of $\phi= 0.5$, $((\sqrt(2)+1/2)\epsilon_d
mN^2/\pi\hbar^2 N_e<<1)$ since it is the richest
possible situation of this system. For other values, near $\phi=0.25$ for 
instance, either the system is outside the Kondo regime for $ N_e= 4n+1,4n+3 $, or,
for other occupations, the situation is conceptually identical to the cases we 
discuss for $\phi\sim0.5$.
The situation  $N_e=4n+1$ and $\phi\sim0.5$ is not
interesting for a small ring. In this case, as soon as the electron goes into the dot
there is no electron left in the ring capable of being excited into a 
unoccupied state with a small cost of energy to which the spin at the dot can 
have a Kondo-like coupling. For the other three possibilities and $ T<T_k$, 
when there is an electron in the dot, the system is in the Kondo regime  and 
the PC has a smoother dependence upon the gate potential than 
it would have for $T>T_k$. See, for example, the charge into the QD and the PC for a small ring shown in Fig. 1c,1d ($N_e=6$ correspond to the Kondo regime and $N_e=7$ not, as discussed below). 
This occurs because in these cases, when the ring electron drops into 
the dot, the wave function  describing an empty dot participates in the Kondo
ground state with a weight that depends upon the Kondo temperature
\cite{12}.
Let us suppose that $N_e=4n+3$, and that the electron that
entered the dot  had a momentum $k_n$ in the
ring. In this
case, the Kondo ground state has a finite weight coming from the state where 
one electron of momentum $k_{-n-1}$ transfers into the $k_{n}$ wave vector, 
creating a vacancy occupied by the electron at the dot, 
antiferromagnetically correlated with the other electron left at
$k_{-n-1}$. This process can take place because the $k_{-n-1}$ and $k_{n}$ 
states differ in energy by an amount less than $T_k$, 
which requires  $\phi\sim0.5$ and also exist for the $4n$ configuration.
However, for the cases $N_e=4n+2$, $(4n)$, the creation 
of the vacancy  at the ring is not necessary because there is
only one (three) electron(s) in the (two) most energetic state(s) that permits 
the formation of a Kondo ground state for any value of $\phi$.  
These results are illustrated in a small ring of 6 atoms laterally
connected to a dot (seven sites in total) with $\phi=0.45$. For this ring size and value of $\phi$ the $N_e=4n+3$ case, $( N_e=7)$, is already outside the Kondo regime, while 
the $N_e=(4n+2)$, $N_e=6$, case is inside it. This is shown in Fig. 2b where we present the spin-spin correlation between the electron at the dot and its first neighbor at the ring for these two cases. 
\begin{figure}
\vspace{-3.cm}
\epsfxsize=2.5in 
\epsffile{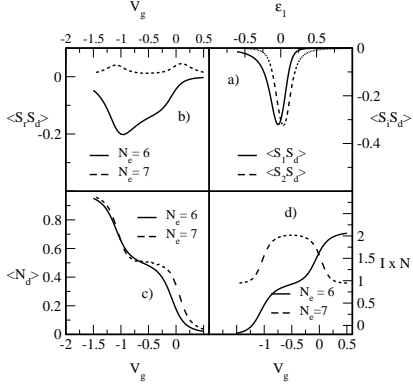}
\epsfysize=2.5in 
\caption{1a)Spin-spin correlation function for a
system composed by two levels $\epsilon_1,\epsilon_2$ and the dot
level $\epsilon_d$  as a function of $\epsilon_1$.
The spin- spin correlation function (1b), the charge in the dot (1c) and the PC(1d)
 as a function of $V_g$, for a dot plus
a six sites ring for $N_e=6$ and $N_e=7$, obtained by exact
diagonalization.}
\label{1} 
\end{figure}

For large systems instead, $\Delta < T_k$, the Kondo effect is present for all
 values of  $N_e$.
In order to pick up the effect of the Kondo correlations in this case,
we study the PC as a function of  the length of the ring in two
situations, $N_e=4n$ and $N_e=4n+3$. The two 
other possible values of $N_e$ do not incorporate any new concept into 
the problem. We take the flux to be $\phi = 0.45$, which according to the 
above discussion, as it is a value in the neighbourhood of $\phi = 0.5$, 
permits us to study for the $N_e=4n+3$ case  the influence of the Kondo effect
on the PC for relatively small rings, $N >100$. In the $\Delta \ll T_k$
regime, our calculations show that when the dot is in 
the Kondo regime, in a region of width $T_k$ in the neighbourhood of the Fermi 
level, the states reduce their conductivity, being non-conducting in the 
immediate vecinity of the Fermi level, as could be expected from the study
of other equivalent systems \cite {8}. 
This conclusion was obtained by analyzing the contribution to the PC, 
$I(w)$, coming from the occupied states. The states with energy below the 
Fermi level by an amount grater than $T_k$ have their participation in the PC 
unaffected by the Kondo spin-spin correlation taking place in the vicinity of 
the dot.

The influence of the Kondo effect in the paired states $k^{\pm}_m={2\pi\over N}
(\pm m+\phi)$ with energies in the neighbourhood of Fermi level $E_f$
produces a change in the PC by an amount proportional to $T_k/(E_fN)$ 
that is negligible in comparison with the contribution proportional to $1/N$, 
coming from the rest of the occupied unpaired states 
(in the semiconductors used to study these phenomena the $E_f $ 
is typically 10mev from the bottom of the conduction band, while $ T_k$ is of the order of 
$10 \mu ev$).  The influence of the Kondo effect is then restricted to the 
non-paired states at the top of the ring occupied states, when an electron 
has already droped into the dot. For $T>T_k$ these states contribute to the 
PC  with a term of the order of $1/N$, that is reduced, and eventually eliminated, 
for $ T < Tk $ due to the localization produced by the Kondo effect. 
This refers us to a similar situation analyzed above when we discussed 
the $\Delta > T_k$ regime.

In Fig. 2 we present  the PC, $I$,  as a function of the gate potential  
for different values of $N$, for a ring of $N_e=4n$ electrons when the dot is 
empty, down to $N_e=4n-2$ when the dot is occupied by two electrons. 
The more remarkable  behaviour here is that as soon as the ring is 
long enough for the system to be in the Kondo regime, the PC varies slowly 
when $V_g$ is reduced, although the ring 
looses an electron that drops into the dot. This variation on $V_g$ disappears completely 
in the thermodynamical limit, when $N= \infty$. 
This can be obtained from the extrapolated value of the current, 
which, as shown in the inset of the figure, satisfies the expression 
$I=I_0+{\alpha\over N^{(1+\beta)}}$, where $I_0$ is the PC corresponding 
to a perfect ring of $4n$ electrons and $\alpha$ and $\beta$ is a constant 
and a positive exponent both depending upon the $V_g$ and $U$. 
We present results as well in the limit $U=\infty$, for a system in  
the $\Delta < T_k$ regime. In this case, for any $ N $, 
the current is independent of the gate potential that, in the absence of the electronic correlation, would 
be strongly dependent upon it. This result coincides with an exact study based 
on a Bethe ansatz procedure \cite{9}, applicable to this system for restricted
 conditions ($U = \infty$; $\phi= 0.5$).
Our calculation shows that, in this limit, their  result is correct for any 
value of $\phi$. We conclude as well that, 
although the dependence of the current on $V_g$ is reduced by the Kondo effect, 
it can not be neglected for the case of a real system with a finite value of $U$ and 
the length of a typical ring. This result is presented in Fig. 2 where the current have been extrapolated to $N=10^4$, for an arbitrary value of $V_g$.   

\begin{figure}
\vspace{-3cm}
\epsfxsize=2.5in
\epsffile{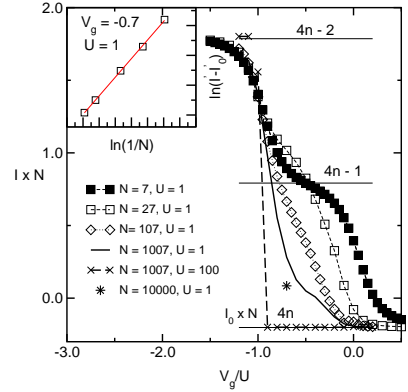}
\epsfysize=2.5in 
\caption{The scaled current $ I^{'}=I$x$N$ as function of $V_g$ for
$N_e=4n$ and different values of  $N$. The inset shows the
scaling of $I^{'}$ with $N$ for $V_g =-0.7$. For the same $V_g$ it is shown
the extrapolated value of $I$ to $N=10^4$. $I_0$ is the PC corresponding 
to a perfect ring of $4n$ electrons.}
\label{2}
\end{figure} 

The role played by the Kondo effect can be understood in the following way. 
When $N_e= 4n$, as soon an electron enters into the dot, there is an unpaired  
electron in the ring capable of providing a finite contribution to the current
that, however, is partially quenched due to the localization of the electronic state 
produced by the Kondo effect. As far as the PC is concerned, this quenching reduces the effect of the charge dropping outside the ring into the dot.

There is a completely different situation when $N_e=4n+3$, as shown in Fig.3. 
In this case there is one unpaired electron contributing to the current on 
the top of the occupied states when the dot is empty. A small ring is
outside the Kondo regime. When the 
highest energetic electron of this system drops  into the dot, 
the current increases because now there are two uncompensated electrons 
contributing. However, for longer rings, when the system is in the Kondo 
regime, this  contribution is  reduced  and the situation changes.
As depicted in Fig. 3, the PC drops from a value corresponding to $N_e = 4n+3$
, towards the value corresponding to a ring of $N_e = 4n$, that is 
reached in the limit $N=\infty$ or
$U=\infty$. Similarly to the case $ N_e= 4n$, 
this behaviour is a consequence of the localization of the 
unpaired states produced by the Kondo effect.
Previous studies that predict 
an independence of the PC on the gate potential for this 
case apparently were not able to establish the essential 
difference between the situations $N_e=4n$ and $N_e=4n+3$.  
The two other cases $N_e=4n+2$ and $N_e=4n+1$ can be analyzed 
following the same ideas developed above without introducing new concepts 
into the discussion. 

\begin{figure}
\vspace{-3cm}
\epsfxsize=2.5in
\epsffile{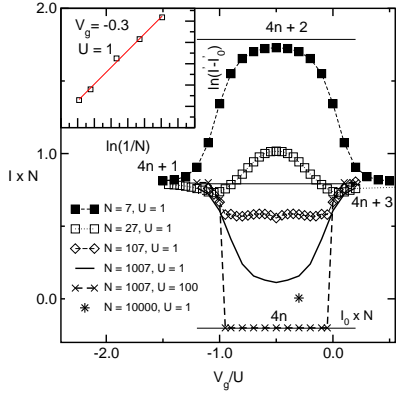}
\epsfysize=2.5in 
\label{3}
\caption{The scaled current $ I^{'}=I$x$N$ as function of $V_g$ for
$N_e=4n+3$ and different values of $N$. The inset shows the
scaling of $I^{'}$ with $N$ for $V_g =-0.3$. For the same $V_g$ it is shown
the extrapolated value of $I$ to $N=10^4$. $I_0$ is the PC corresponding 
to a perfect ring of $4n+3$ electrons.}
\end{figure}

In summary, we have studied the persistent current going along a ring with a 
dot side-coupled to it. Our results show that the system could be in the 
Kondo regime independently of its size, $\Delta {>\atop <} T_k$, and that, due 
to the Kondo spin-spin correlation, the states near the Fermi level 
 are not conducting, similarly to what occurs when a magnetic impurity is 
embeded in a metal. This clarifies the controversy regarding the apparent 
different behaviour of a dot connected to a wire and to a ring. We were able 
to show that the physics of the system is controlled by the unpaired most 
energetic occupied states that in the Kondo regime provides a slow dependence of 
the current with the gate potential for the case $N_e= 4n$ and a 
strong dependency for $N_e=4n+3$.

These properties make the detection of the Kondo effect through the 
measurement of the PC of a ring with a side-coupled dot a very interesting 
experiment.

This work was partially supported by CONICET, UBA (UBACYT X210 ), Fundacion
Antorchas  and Vitae/Antorchas/Andes, grant B-11487/9B003.

\end{document}